\documentclass[pre,twocolumn,twoside,byrevtex,superscriptaddress,floatfix,
showpacs]{revtex4-1}

\usepackage{currfile}
\usepackage{color}

\usepackage{graphicx,epsfig,verbatim,enumerate}
\usepackage{amssymb,amsmath}
\usepackage{ifthen}

\usepackage{longtable}
\usepackage{algpseudocode} 

\usepackage{mathtools}

\newboolean{twocolswitch}

\newcommand{\sindex}[1]{}
\newcommand{\nindex}[1]{}

\newcommand{\www}[1]{\url{#1}}

\newcommand{\eps}{\varepsilon}

\usepackage{lettrine}

\renewcommand{\vec}[1]{\mathbf{#1}}

\renewcommand{\i}{\mathbf{i}}

\setboolean{twocolswitch}{true}

\begin{document}

\title{\protect
Continuum rich-get-richer processes: \\
Mean field analysis with an application to firm size
}

\author{
\firstname{David Rushing}
\surname{Dewhurst}
}
\email{david.dewhurst@uvm.edu}

\affiliation{The Mitre Corporation, McLean, VA}
\affiliation{Department of Mathematics \& Statistics,
  Vermont Complex Systems Center,
  Computational Story Lab,
  \& the Vermont Advanced Computing Core,
  The University of Vermont,
  Burlington, VT 05401.}

\author{
	\firstname{Christopher M.}
	\surname{Danforth}
}
\email{chris.danforth@uvm.edu}

\affiliation{Department of Mathematics \& Statistics,
  Vermont Complex Systems Center,
  Computational Story Lab,
  \& the Vermont Advanced Computing Core,
  The University of Vermont,
  Burlington, VT 05401.}

\author{
\firstname{Peter Sheridan}
\surname{Dodds}
}
\email{peter.dodds@uvm.edu}

\affiliation{Department of Mathematics \& Statistics,
  Vermont Complex Systems Center,
  Computational Story Lab,
  \& the Vermont Advanced Computing Core,
  The University of Vermont,
  Burlington, VT 05401.}

\date{\today}

\begin{abstract}
  \protect
  Classical rich-get-richer models have found much success in being able to broadly reproduce the statistics and dynamics of diverse real complex systems.
These rich-get-richer models are based on classical urn models and unfold step-by-step in discrete time. Here, we consider a natural variation acting on a temporal continuum in the form of a partial differential equation (PDE). 
We first show that the continuum version of Herbert Simon's canonical preferential attachment model exhibits an identical size distribution. 
In relaxing Simon's assumption of a linear growth mechanism, we consider the case of an arbitrary growth kernel and find the general solution to the resultant PDE. 
We then extend the PDE to multiple spatial dimensions, again determining the general solution.
We then relax the zero-diffusion assumption and find an envelope of solutions to the general
model in the presense of small fluctuations. 
Finally, we apply the model to size and wealth distributions of firms. 
We obtain power law scaling for both to be concordant with simulations as well as observational data, 
providing a parsimonious theoretical explanation for these phenomena.
\end{abstract}


 \maketitle


\section{Introduction}
\label{sec:introduction}

In 1955, Herbert Simon described a general version of a rich-get-richer process that generates power-law size distributions 
$P(s) \sim s^{-\gamma}$ with scaling exponent 
$\gamma >2$~\cite{simon1955class}. 
Simon's process was adapted by Price to capture the statistics of growing networks, and was later paralleled by the Barab\'{a}si-Albert model which introduced scale-free networks~\cite{barabasi1999emergence}. 
Simon's model efficiently captures the statistical properties of a wide variety of real-world phenomena, such as the linking dynamics of the Web~\cite{bornholdt2000topological} and the growth of software distributions~\cite{maillart2008empirical}. Recently, the present authors and others have shown Simon's model also exhibits a potentially pronounced first-mover advantage and that this feature may be consistent with the growth of real systems~\cite{dodds2017a}.

In Sec.\ \ref{sec:model-and-analysis}, we first realize Simon's model in a continuum setting and describe its dynamics for a number of growth kernels.
Secs.\ \ref{sec:discrete-to-cont} and \ref{sec:general-model} describe the continuum version of the model and formulate its analytical solution.
In Sec.\ \ref{sec:examples} we determine analytically how the size distribution generated by the process is dependent on the growth kernel, and can be proportional not only to any power law distribution with finite mean ($\gamma >2$), but also specific instances of the extreme value distribution, while in Sec.\ \ref{sec:multidim} we analyze the model's behavior when extended to many dimensions. 
We apply the model to the dynamics of a market economy in Sec.\ \ref{sec:econ}, showing that the power law distribution of firms observed empirically and in simulation can be derived from first principles of microeconomic theory with a minimum of assumptions.
\footnote{All data and code to recreate figures is available on the lead author's website: \protect\url{https://github.com/daviddewhurst/continuum-preferential-attachment}}

\section{Model and analysis}
\label{sec:model-and-analysis}

We describe Simon's discrete model by means of an economic example.
Suppose an individual creates a new firm in some product space with themselves as the sole employee.
An individual that enters the product space at time-step $t$ must choose between starting a new firm themselves with probability $\rho$, and choosing to join an existing firm with probability $1-\rho$ from one of the existing firms, with the likelihood of choosing any particular firm from which to purchase proportional to the number of  employees $k$.
We will denote the number of firms of size $k$ at time $t$ by $N_{k, t}$. 
The general discrete model thus takes the form of the recurrence relation \cite{simon1955class}
\begin{equation}\label{eq:simon-discrete}
\langle N_{k,t+1} - N_{k,t}\rangle
=
(1-\rho)\Big( -\frac{k}{t}N_{k,t}+\frac{k-1}{t}N_{k-1,t}\Big)
\end{equation}
where we formalize $\rho$ as an innovation probability. 
The solution to (\ref{eq:simon-discrete}) scales as 
\begin{equation}\label{eq:simon-discrete-asymptotic-sol}
N_{k,t} \sim tk^{-\gamma},
\end{equation}
with $\gamma = 1 + \frac{1}{1 - \rho}$. 
When $\rho \rightarrow 0$, the size exponent $\gamma \rightarrow 2$, so that the distribution thus obtained borders on infinite mean. 
Zipf's law for rank-frequency distributions, written $s_r \propto r^{-\alpha}$, is recovered from (\ref{eq:simon-discrete-asymptotic-sol}) by setting the Zipf exponent $\alpha = \frac{1}{\gamma - 1} = 1 - \rho$ \cite{simon1955class}. 
The corresponding equation for the size of the $n$-th arriving group $S_{n,t}$ is then given by \cite{dodds2017a}
\begin{equation}
S_{n, t} =
\begin{cases}
\begin{alignedat}{2}
&\frac{1}{\Gamma(2 - \rho)}\left[\frac{1}{t}\right]^{-(1 - \rho)} &&\quad \text{if }n = 1\\
&\rho^{1 - \rho}\left[ \frac{n - 1}{t} \right]^{-(1 - \rho)} &&\quad \text{if }n \geq 2
\end{alignedat}
\end{cases}
\end{equation}
\subsection{From discrete to continuous}\label{sec:discrete-to-cont}
Though the discrete model accurately models the size distribution resulting from many real-world processes, it has a number of shortcomings when applied to economic situations. 
Models that exploit mathematical properties of the preferences of a representative agent often perform poorly in the task of explaining economic phenomena such as inequality and skewed wealth distributions \cite{epstein2006remarks,sen1977rational}. 
Abstraction of these considerations is thus desirable in order to account for idiosyncrasies present at the individual consumer level; the resulting model is a mean-field equation and better describes macro behavior, analogous to the use of deterministic equations of statistical mechanics to describe stochastic interactions among many particles. 
Moving from discrete to continuous time is sensible as it corresponds better with our notion of reality (people do not make decisions in synchrony at each tick of a universal clock).
From a mathematical viewpoint, the resulting equation will be more easily analyzed as a partial differential equation instead of a coupled differential-difference equation. 
Further, where Simon's model assumes that agents aggregate to firms with growth kernel $r(x) = x$, we drop this assumption and write the growth kernel as some function $r(x)$ to allow for generalization of choice \cite{krapivsky2001a}.
Finally, we allow the innovation rate $\rho$ to vary in time as $\rho(t)$, which may more realistically capture the process of technological innovation inherent in the present economic system. 
(We note that Simon considered a time-varying innovation probability in \cite{simon1955class}.)
\medskip

\noindent
While other authors have considered models that provided important contributions to the 
understanding of rich-get-richer processes and are {\it prima facie} similar to ours 
\cite{saldana2007continuum, krapivsky2001a}, our model differs substantially from those previously created. 
Previous continuum models have focused solely on networks, while ours is intentionally more general, allowing
us to construct parsimonious models of economic phenomena, for example. 
In addition, these previous models did not treat the most general problem of arbitrary growth kernel 
$r(x)$ and innovation rate $\rho(t)$ in the manner considered here.
In addition, the continuum formulation extended to an arbitrary (finite) number of 
dimensions is entirely novel as far as we are 
aware; this formalism can be used in modeling interacting preferential attachment processes in the fields
of biology, economics, or sociology.

\subsection{General asymptotic model}\label{sec:general-model}
Applying the above adjustments to Simon's discrete model, Eq.\ \ref{eq:simon-discrete} becomes a boundary value problem for the function determining the intertemporal distribution of firms of size $x$, written $f(x,t)$:
\begin{equation}\label{eq:pde-gen}
\frac{\partial f}{\partial t} = -\frac{1-\rho(t)}{t}\frac{\partial}{\partial x}[r(x)f]
\end{equation} 
with the semi-infinite boundary condition $\lim_{x\rightarrow \infty} f(x,t) = 0$. 
(We treat only the case of asymptotic solutions; the question of differing initial distributions of firms is not considered.)
We first consider $\rho_{\infty}$ as a long-run constant innovation rate satisfying $\lim_{t \rightarrow \infty}\frac{\rho(t)}{\rho_{\infty}} = 1$ and solve Eq.\ \ref{eq:pde-gen} by separation of variables. 
Setting $f(x,t) = X(x)T(t)$, we solve the equations: 
\begin{align}\label{eq:simon-continuous-general-odes}
\frac{dT}{dt} &\simeq \frac{\lambda}{1-\rho_{\infty}}\frac{1-\rho(t)}{t}T(t) \\
 \frac{dX}{dx} &=-\Big[\frac{\lambda}{(1-\rho_{\infty})r(x)} + \frac{r'(x)}{r(x)} \Big]X(x), 
\end{align}
where $\lambda$ is a constant of separation. 
The general solution to (\ref{eq:pde-gen}) is thus 
\begin{equation}\label{eq:simon-continuous-general-sol}
f(x,t) = \frac{c}{r(x)}\exp\Big[ \frac{\lambda}{1-\rho_{\infty}}g(x,t) \Big] 
\end{equation}
where $g(x,t) = \int^t\frac{1-\rho(t')}{t'}dt' - \int^x\frac{dx'}{r(x')}$. 
\medskip

\noindent
Dropping the assumption that $\rho(t) \rightarrow \rho_{\infty}$, we solve (\ref{eq:pde-gen}) in all generality using the method of characteristics.
We write the Lagrange-Charpit equations that describe its solution on the characteristic curves as 
\begin{equation}\label{eq:characteristics}
dt = \frac{t\ dx}{(1-\rho(t))r(x)} = -\frac{t\ df}{(1-\rho(t))r'(x)f(x,t)} 
\end{equation}
and solve the resulting equations
\begin{align}\label{eq:lag-charp-odes}
	\frac{dx}{dt} &= \frac{1-\rho(t)}{t}r(x) \\
	\frac{df}{dx} &= -\frac{r'(x)}{r(x)}f(x,t)
\end{align}
The solution to the first is given implicitly by $\int \frac{dx}{r(x)} + A = \int \frac{1-\rho(t)}{t}dt$, while the solution to the second is $f(x,t) = \frac{B}{r(x)}$. 
Letting $B = F(A)$ gives the firm density
\begin{equation}\label{eq:pde-gen-sol}
f(x,t) = \frac{1}{r(x)}F\Big(\int\frac{1-\rho(t)}{t}dt - \int \frac{dx}{r(x)} \Big)
\end{equation} 
We see that Eq.\ \ref{eq:simon-continuous-general-sol} has the same form as Eq.\ \ref{eq:pde-gen-sol}, with $F(\cdot) = \exp (\cdot)$. 
\subsection{Asymptotics for example growth kernels}\label{sec:examples}
\begin{figure}[tp!]
  \centering
  \includegraphics[width=\columnwidth]{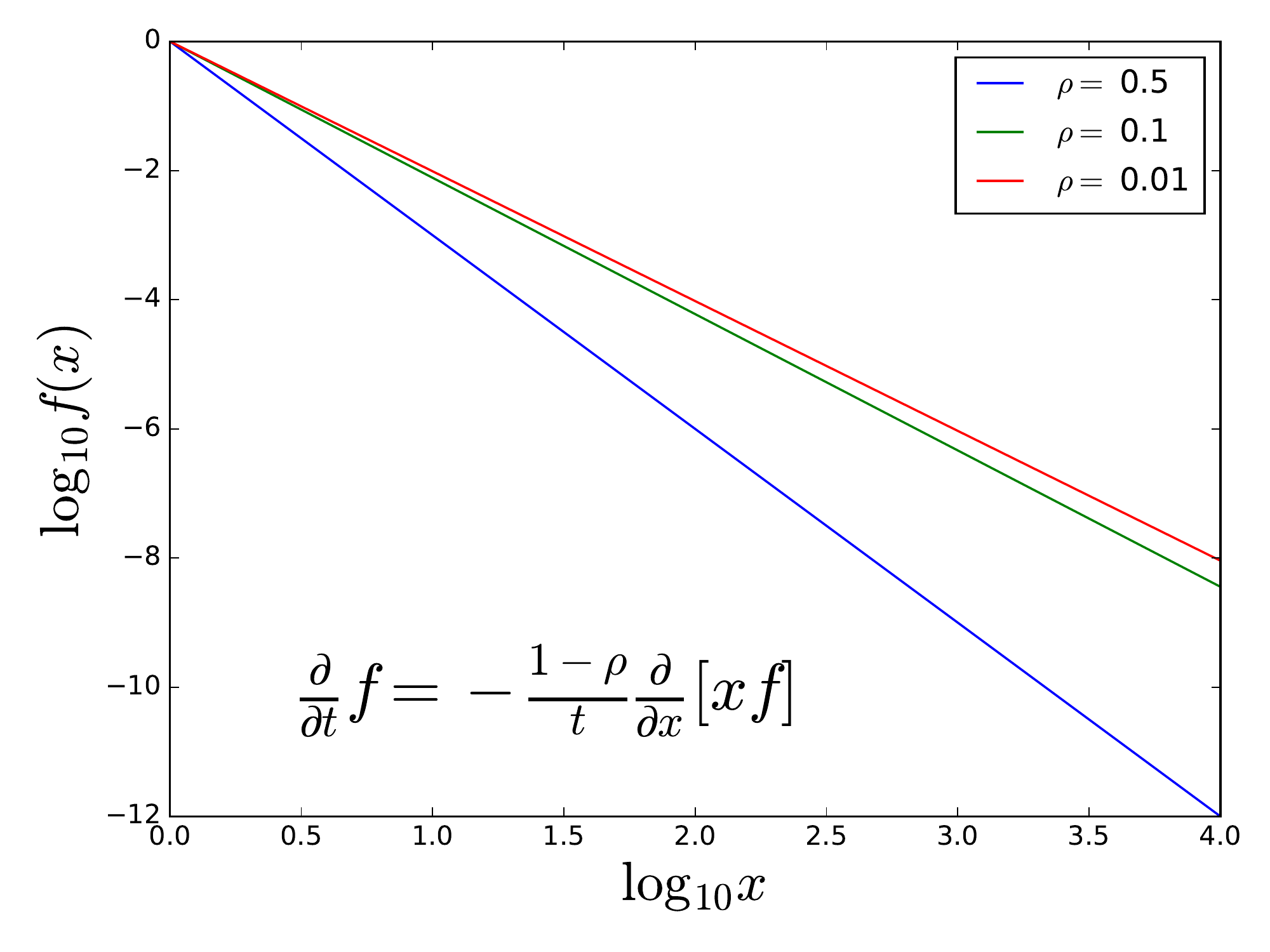}
  \caption{ 
  Solutions to Eq. \ref{eq:pde-gen} with growth kernel $r(x) = x$ and innovation rate $\rho(t) = \rho_{\infty}$.
  }
  \label{fig:solution-plots}
\end{figure}
\begin{figure}[tp!]
  \centering
  \includegraphics[width=\columnwidth]{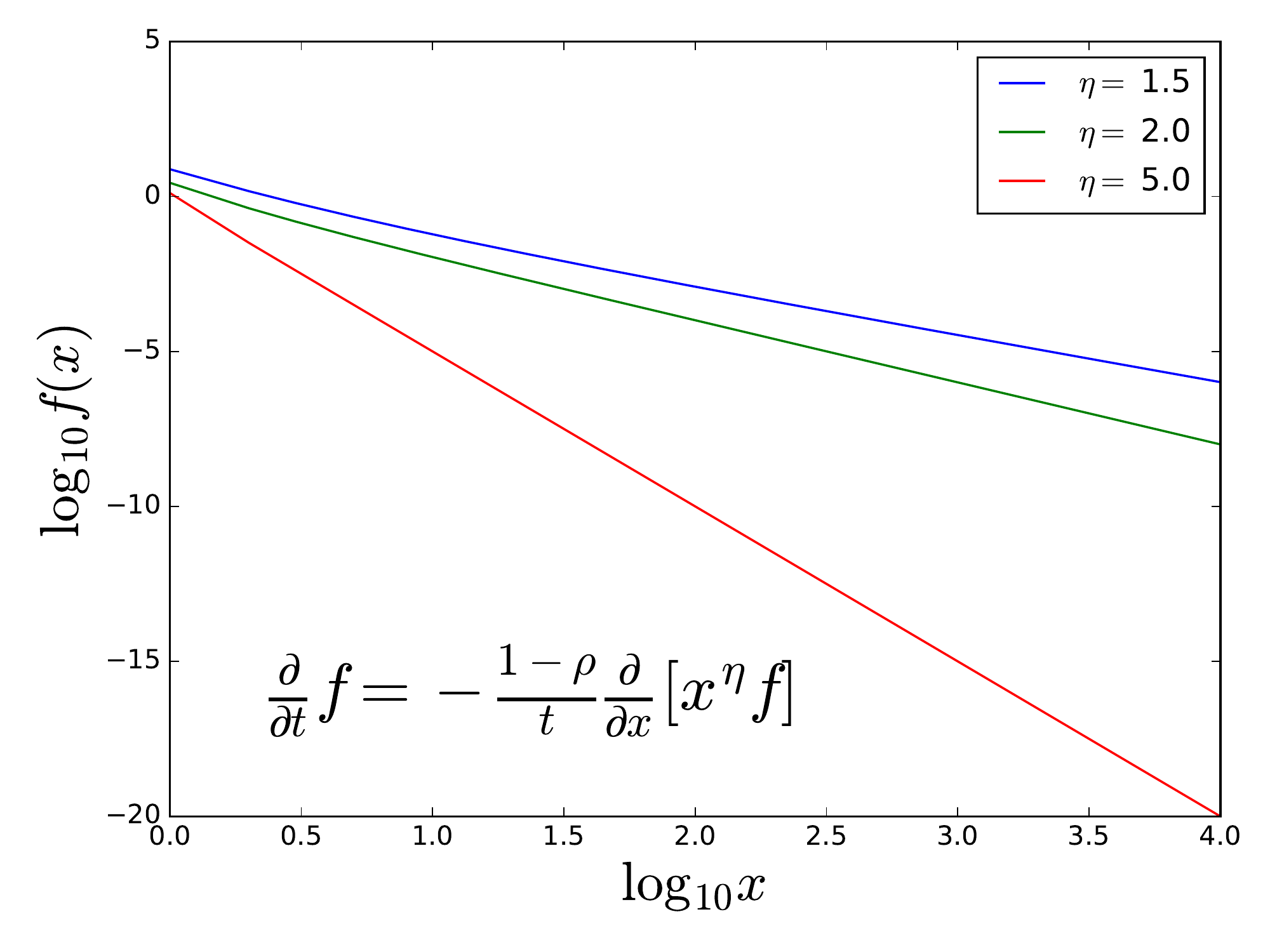}
  \caption{ 
  Solutions to Eq.\ \ref{eq:pde-gen} with growth kernel $r(x) = x^{\eta}$ for $\eta > 1$ and innovation rate $\rho(t) = \rho_{\infty}$.
  }
  \label{fig:solution-plots2}
\end{figure}

We wish to characterize the long-run behavior of equation (\ref{eq:simon-continuous-general-sol}). 
Recovering the original Simon model is possible by setting $r(x) = x$ and letting the innovation rate remain constant at $\rho_{\infty}$.
Equation (\ref{eq:simon-continuous-general-sol}) then becomes
\begin{align}\label{eq:recovery-simon}
f(x,t) &\propto \frac{1}{x}\exp \Big[ \ln t -\frac{\lambda}{1-\rho_{\infty}}\int^x\frac{dx'}{x'} \Big] \notag \\
&= tx^{-(1+\frac{\lambda}{1-\rho_{\infty}})}
\end{align}
with $\lambda \rightarrow 1$ in the long-run time limit.
Figure \ref{fig:solution-plots} shows solutions of Eq. \ref{eq:pde-gen} with the Simon growth kernel $r(x) = x$ and a constant innovation rates $\rho \in \{0.01, 0.1, 0.5\}$
Note that these solutions are pure power laws and thus are linear in log-log space. 
The exponent $\gamma = 1+\frac{1}{1-\rho_{\infty}}$ is the expression found by Simon in \cite{simon1955class}.
Any discretization of this continuum process will have a size-rank distribution $S(r) \propto r^{-\alpha}$ with Zipf exponent $\alpha = \frac{1}{\gamma - 1} = 1- \rho_{\infty}$; the dynamics of the Simon process are thus completely recovered in this case.
\medskip

\noindent
For an affine growth kernel $r(x) = a_0+a_1x$, the solution is similar:
\begin{align}\label{eq:recovery-genlinear}
f(x,t) &\propto \frac{1}{a_0+a_1x}\exp\Big[ \ln t-\frac{\lambda}{1-\rho_{\infty}}\int^x\frac{dx'}{a_0+a_1x'} \Big] \notag \\
&= t(a_0+a_1x)^{-1-\frac{\lambda}{a_1(1-\rho_{\infty})}} \notag \\
&\sim tx^{-(1+\frac{1}{a_1(1-\rho_{\infty})})},
\end{align}
again in the long-run time limit.
A general linear growth factor can thus be chosen to result in a power law distribution with any $\gamma >1$; as $a_1$ grows large, $\frac{1}{a_1(1-\rho_{\infty})} \rightarrow 0$.
\medskip

\noindent
Considering a monomial power growth factor $r(x) = x^{\eta}$ ($\eta \neq 1$), we obtain
\begin{align}\label{eq:recovery-power}
f(x,t) &\propto \frac{1}{x^{\eta}}\exp \Big[ \ln t - \frac{1}{1-\rho}\int^x\frac{dx'}{x^{'\eta}} \Big] \notag \\
&= tx^{-\eta}\exp\Big[ -\frac{x^{-(\eta -1)}}{(1-\rho)[-(\eta-1)]} \Big].
\end{align}
Equation (\ref{eq:recovery-power}) is proportional to a Fr\'{e}chet distribution for $\eta > 1$ or Weibull distribution for $0 < \eta <1$. 
Krapivsky \textit{et al.}\ found a similar result for $n_k$, the number of nodes of degree $k$, in growing random networks \cite{krapivsky2000connectivity}. \\
~\\
Figure \ref{fig:solution-plots2} shows solutions of Eq. \ref{eq:pde-gen} with $r(x) = x^{\eta}$ for $\eta >1$. 
 
\subsection{Preferential attachment in many dimensions}\label{sec:multidim}
We now extend the model, Eq.\ (\ref{eq:pde-gen}), to multiple dimensions.
The general model reads
\begin{equation}
	\begin{aligned}
		\frac{\partial f}{\partial t} &= -\frac{1 - \rho(t)}{t}\mathcal{L}_{\text{div}}
		(r(x_1,...,x_N), f)\\
		&\qquad\qquad + \mathcal{L}_{\text{fluc}}(D(x_1,...,x_N)f),
	\end{aligned}
\end{equation}
where $r(x_1,...,x_N)$ is a general growth kernel, $\mathcal{L}_{\text{div}}$ is a linear differential operator containing 
a divergence term, the generator of the rich-get-richer process, and $\mathcal{L}_{\text{fluc}}$ is an operator
containing information on the random fluctuations of $f$.
We describe several models and solve one of them analytically.
\medskip

\noindent
The most straightforward generalization of Eq.\ \ref{eq:pde-gen} holds constant the assumption of zero fluctuations
and has generator given by the divergence of the vector field 
$\vec{E} = \sum_{i=1}^N (r_k(x_k) f) \vec{e}_k$, with $\vec{e}_k$ the standard orthonormal basis for the particular vector
space under study.
In the case of cartesian coordinates, the generator becomes 
$$
\mathcal{L_{\text{div}}} = 
\sum_{i=1}^N \frac{\partial}{\partial x_k}(r_k(x_k)f).
$$ 
This case has an analytical solution given below in 
Sec.\ \ref{sec:separable-multi} that directly parallels the results given in Sec.\ \ref{sec:general-model}.
When $r_k = r_k(x_1,...,x_N)$, the resulting equation is not separable and resists analytical solution. 
\medskip

\noindent
A reasonable generalization relaxes the zero-diffusion assumption to study the process under the 
influence of small perturbations.
Assuming small random fluctuations gives a time-dependent Fokker-Planck equation for $f$ which we now describe. 
Letting $\vec{D} = \sum_{i,j}D_{ij}(x_1,...,x_N)$ be the covariance matrix, we have (in Cartesian 
coordinates)
\begin{equation}\label{eq:pref-att-diff}
	\frac{\partial f}{\partial t} = -\frac{1 - \rho(t)}{t}\nabla \cdot \vec{E} 
+ H[\vec{D}f],
\end{equation}
where $H= \sum_{i,j}\frac{\partial^2}{\partial x_j \partial x_i}$ is the diffusion operator.
This equation is also, in general, not possible to solve analytically.
However, an envelope of solutions is available in any small time interval, as shown in Section 
\ref{sec:diffusion}. 
 
\subsubsection{Separable growth kernels $r_k$}\label{sec:separable-multi}
Maintaining the assumption of a steady-state constant innovation rate $\rho_{\infty}$, the equation governing the distribution of firms as a function of time and $N$ spatial variables $x_1,...,x_N$ is
\begin{equation}\label{eq:multidim}
\frac{\partial f}{\partial t} 
= 
-\frac{1-\rho(t)}{t}\sum_{k=1}^N \frac{\partial}{\partial x_k}(r_k f),
\end{equation}
where $r_k = r_k(x_k)$. 
This equation is again separable with solution given by
\begin{equation*}
f(x_1,...x_N,t) = T(t)\prod_{k=1}^NX_k(x_k).
\end{equation*}
Substituting the above into Eq.\ (\ref{eq:multidim}) gives 
\begin{equation}\label{eq:multidim-rewrite}
\begin{aligned}
\frac{\partial[T(t)\prod_1^NX_k]}{\partial t} 
= 
-\frac{(1-\rho_{\infty})(1-\rho(t))}{(1-\rho_{\infty})t} \\
\times\sum_{k=1}^N \frac{\partial}{\partial x_k}\Big(r_k T(t)\prod_{j=1}^NX_j\Big),
\end{aligned}
\end{equation}
which, after rearranging and differentiating, becomes
\begin{equation}\label{eq:multidim-diff}
\begin{aligned}
\frac{1-\rho_{\infty}}{1-\rho(t)}&\frac{t}{T(t)}\frac{dT}{dt} \prod_{k=1}^NX_k = -(1-\rho_{\infty}) \\
&\times \sum_{k=1}^N \Big( r_k\frac{dX_k}{dx_k}\prod_{j\neq k}X_j+\frac{dr_k}{dx_k}\prod_{j=1}^NX_j \Big).
\end{aligned}
\end{equation}
Dividing through by the term $\prod_{k=1}^NX_k$, we find
\begin{equation}\label{eq:multidim-divbyprod}
\begin{aligned}
\frac{dT}{dt}\frac{(1-\rho_{\infty})t}{(1-\rho(t))T(t)} &= -(1-\rho_{\infty}) \\
&\times\sum_{k=1}^N\Big( r_k\frac{dX_k}{dx_k}X_k^{-1} + \frac{dr_k}{dx_k} \Big), 
\end{aligned}
\end{equation}
which can be separated into an uncoupled system of $N+1$ ODEs with structures identical to those solved in Eqs.\ (\ref{eq:simon-continuous-general-odes}):
\begin{align}\label{eq:ode-multi}
\frac{dT}{dt} &= \frac{\lambda}{1-\rho_{\infty}}\frac{1-\rho(t)}{t}T(t), \\
\frac{dX_k}{dx_k} &= -\left( \frac{\lambda_k}{(1-\rho_{\infty})r_k} + \frac{r'_k}{r_k} \right)X_k \text{ for } k = 1,...,N,
\end{align}
where $\lambda = \sum_{k=1}^N \lambda_k$ are the coefficients of separation. 
The general solution of Eq.\ (\ref{eq:multidim}) is thus
\begin{equation}\label{eq:multidim-sol}
\begin{aligned}
f(x_1,...x_N,t)\propto \frac{c}{r}\exp&\bigg[\frac{1}{1-\rho_{\infty}}\Big(\int^t\frac{1-\rho(t')}{t'}dt' \\ &-\sum_{k=1}^N \int^{x_k}\frac{dx_k'}{r_k(x_k')}\Big) \bigg],
\end{aligned}
\end{equation}
with $c = c_t\prod_k c_k$ and $r = \prod_k r_k(x_k)$.

\subsubsection{The case of nonzero diffusion}\label{sec:diffusion}
Let $\Delta t$ be some small time interval and pick $t_0 < t_1$ so that $t_1 - t_0 < \Delta t$ 
defines some time window of interest.
Define $\tau(t) = \frac{(1 - \rho_{\infty})t}{1 - \rho(t)}$. Fixing $\tau_0 = \tau(t_0)$ and $\tau_1 =
\tau(t_1)$ and substituting $\tau_j,\ j=1,2$ for $\tau$ in Eq.\ \ref{eq:pref-att-diff} define 
solutions $f_0(x_1,...,x_n,t)$ and $f_1(x_1,...,x_n,t)$ whose generators are time-independent and 
whose spatial averages form an envelope for the spatial average of 
the solution of Eq.\ \ref{eq:pref-att-diff} as 
$f_1(t) \leq f(t) \leq f_0(t)$ for $t \in [t_0, t_1]$.
We will solve for the upper envelope solution $f_0$ explicitly; to solve for $f_1$ one proceeds identically.
Again assuming a product solution of the form $f_0(x_1,...,x_n,t) = T_0(t) \prod_{k=1}^n X_k(x_k)$
and supposing that the diffusion matrix is $\vec{D} = \frac{\sigma^2}{2}\vec{I} = D\vec{I}$, 
substitution in Eq.\ \ref{eq:pref-att-diff} gives  
\begin{equation}
	\begin{aligned}
		&\frac{\partial [T_0(t)\prod_{i=1}^n X_i]}{\partial t} = -\tau_0(1-\rho_{\infty})\\
		&\qquad\quad\times\sum_{k=1}^N \frac{\partial}{\partial x_k}
		\left(r_k T_0(t)\prod_{j=1}^NX_j\right)\\
		&\quad\qquad +D\sum_{k=1}^n \frac{\partial^2}{\partial x_k^2}
		\left(T_0(t)\prod_{j=1}^n X_j \right).
	\end{aligned}
\end{equation}
Following a similar derivation to Eqs. \ref{eq:multidim-rewrite} - \ref{eq:ode-multi}, one arrives 
at the system of ODEs
\begin{align}
	\frac{dT_0}{dt} &= \frac{\lambda}{\tau_0}T_0(t) \\
	\tau_0 D \frac{d^2X_k}{dx_k^2} &+ r_k\frac{dX_k}{dx_k} 
	+ \left[\frac{\lambda_k}{1 - \rho_{\infty}} +r_k' \right]X_k = 0 
\end{align}
where again $\sum_{k=1}^n\lambda_k = \lambda$.
The time solution is now given by $T_0(t) = \exp\left( \frac{\lambda}{\tau_0}t \right)$
(recall that this is defined only over $t \in [t_0,t_1]$), while the spatial solutions 
are much more intricate than those given in Eq.\ \ref{eq:multidim-sol}.
Where $r_k(x_k) = 1$ the solution is given in terms of sines and cosines and it is seen that 
Eq. \ref{eq:pref-att-diff} simply becomes the heat equation; there is no preferential attachment 
process here. 
The case of classical preferential attachment is given by $r_k(x_k) = x_k$, whereupon the spatial 
equations take the form 
\begin{equation}\label{eq:hermite-type}
	\frac{d^2X}{dx^2} - c_1 x\frac{dX}{dx} + c_2X = 0,
\end{equation}
where we have set $X = X_k$ for clarity and defined the constants $c_1 = -\frac{1}{\tau_0 D}$ 
and $c_1 = \frac{1}{\tau_0 D}\left( \frac{\lambda}{1 - \rho_{\infty}} - 1\right)$.
This equation is of Hermite type and its solution can be expressed analytically in terms of the confluent
hypergeometric function. 
This does not provide elucidation of the resultant distribution, however; we derive the solution of 
Eq.\ \ref{eq:hermite-type} in frequency space presently.
Defining the Fourier transform by $F(\omega) = F[f](\omega) = \int_{-\infty}^{\infty}f(x) e^{\i \omega x} dx$, 
transforming Eq. \ref{eq:hermite-type} results in the frequency-space differential equation
\begin{equation}\label{eq:freq-ode}
	\frac{\partial F}{\partial \omega} = \frac{1}{c_1 \omega}(\omega^2 - c_1 - c_2)F(\omega).
\end{equation}
The asymptotic solution to Eq.\ \ref{eq:freq-ode} in frequency space is thus (replacing $F$ by $F_k$)
\begin{equation}
	F_k(\omega) \simeq
	\omega^{-\left(2 - \frac{\lambda_k}{1 - \rho_{\infty}}\right)}e^{-\frac{\omega^2}{2\tau_0 D}}.
\end{equation}
We note the decomposition of $F_k(\omega)$ as a product (in frequency space) of a pure diffusion part 
and a preferential attachment (power law) frequency decay; the corresponding time-valued function 
is a convolution of a diffusion process and the preferential attachment process.
\medskip

\noindent
Setting $L^{\dagger} = -\frac{1 - \rho(t)}{t}\mathcal{L}_{\text{div}} 
+ \frac{\sigma^2}{2}\mathcal{L}_{\text{fluc}}$, we can write Eq.\ \ref{eq:pref-att-diff} as 
$\frac{d}{dt}f = L^{\dagger}f$ when the diffusion is uncorrelated (the diffusion matrix is a
multiple of the identity). 
Another way of characteristing solutions to this equation is to solve for a stochastic process 
that generates the equivalent backward solutions; we search for solutions of $-\frac{d}{dt}f = Lf$. 
The corresponding PDE is given by 
\begin{equation}
	-\frac{\partial f}{\partial t} = \frac{1 - \rho(t)}{t}r(x) \frac{\partial f}{\partial x}
	 + \frac{\sigma^2}{2}\frac{\partial^2f}{\partial x^2},
\end{equation}
defined for $t \in [t_0, T]$ with final condition $f(x,T) = \phi(x)$.
(In dimensions higher than one the extension is clear.)
By the Feynman-Kac formula, the solution to this equation is given by 
\begin{equation}
	f(x,t) = \left\langle \phi(X_T) | X_t = x \right \rangle,	
\end{equation}
where the stochastic process $X_t$ is defined by 
\begin{equation}
dX_t = \frac{1 - \rho(t)}{t}r(X_t)\ dt + \sigma\ dW_t.
\end{equation}
Thus analysis of the associated It\^{o} SDE yields yet another 
method by which behavior of the continuum rich-get-richer process can be analyzed.

\begin{figure}[tp!]
  \centering
  \includegraphics[width=\columnwidth]{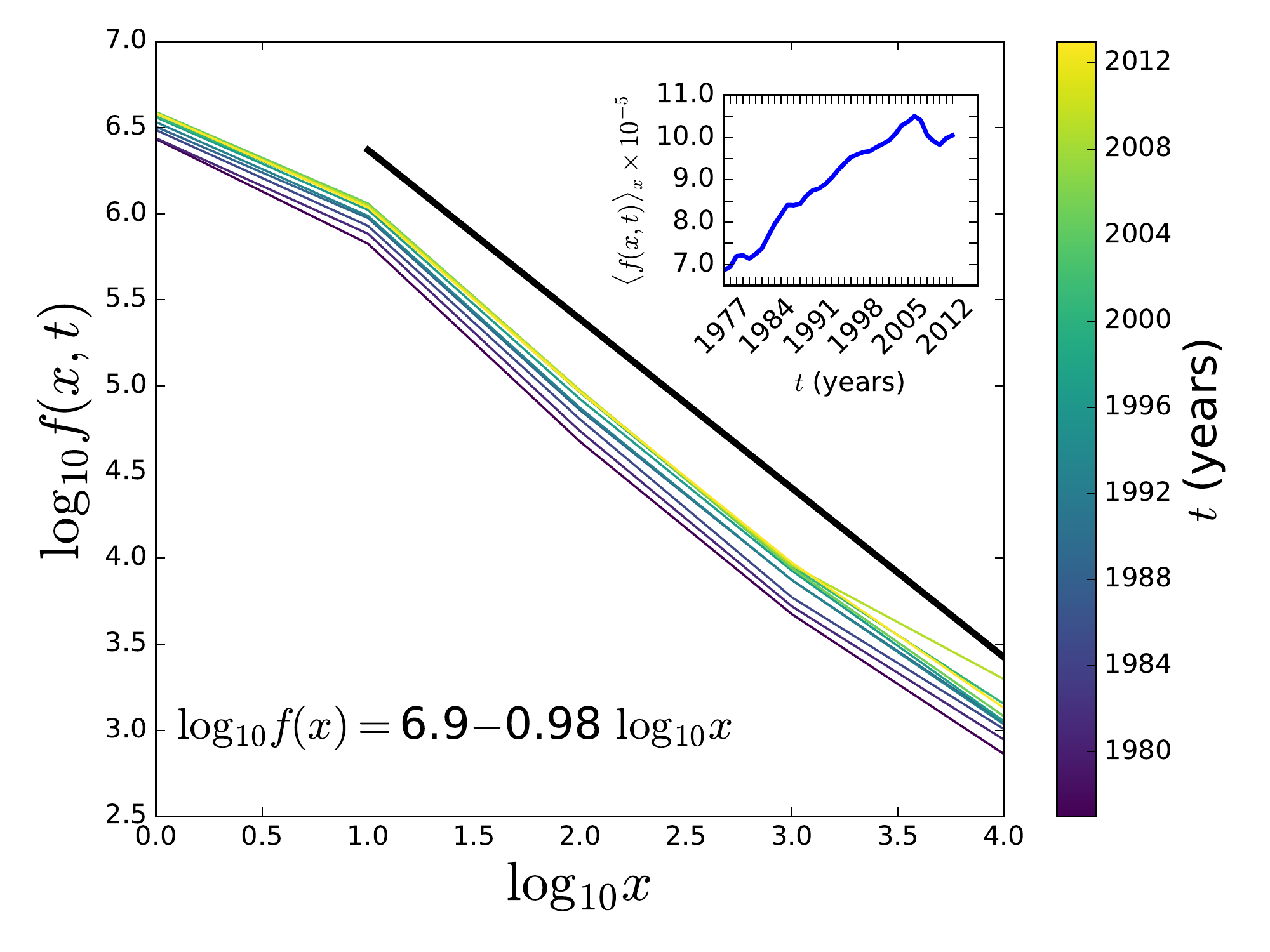}
  \caption{
    \textbf{Power law size distribution of firms.}
Distribution of firm sizes by employment from 1977 to 2013, obtained from the U.S. Census Bureau on August 28, 2016 \cite{firm-size-data}.
Note that U.S. firms exhibit constant returns to scale, so this implies a power-law distribution (with identical exponent) in firm income. 
    This result was famously publicized by Axtell in \cite{axtell2001zipf}. 
The wide binning in the figure is due to the lack of granularity in publicly available U.S. Census data on firms with more than $10^4$ employees.
The inset displays $\langle f(x, t) \rangle_x$ over the entire date range. 
These data exhibit linear scaling, as predicted by Eq.\ \ref{eq:revenue-sol}.
}
  \label{fig:size-powerlaw}
\end{figure}

\section{Application to market structure}\label{sec:econ}
\subsection{Firm size}
We demonstrate the applicability of our results with a microeconomic analysis of firm revenues.
Consider a small time period $\Delta t$ in which consumers enter a market to purchase an item priced at $p$.
During this time period, we assume each consumer purchases only one item.  
With probability $\rho$ (likely quite small) a consumer will choose to start their own firm; with probability $1-\rho$ they will choose to buy the product from an existing firm. 
Consumers chose a firm from which to buy in proportion to the advertising level of the firm, which is itself proportional to the firm's revenue $R$. 
Revenue is given by the equation $R(q) = p(q)\cdot q$, where $q$ is the quantity of the product sold. 
Firms are price takers, with market price set at $p$, so that revenue is $R(q) = pq + c$ for $c$ some constant. 
Applying the above model, the mean-field equation for this process is 
\begin{equation}\label{eq:revenue-proc}
\frac{\partial f(x,t)}{\partial t} = -\frac{1-\rho}{t}\frac{\partial}{\partial x}[(px + c)f(x,t)],
\end{equation}
where the substitution $q = x$ comes from the restriction that each consumer purchases only one product during the small time interval of study. 
The solution to this equation is given by 
\begin{equation}\label{eq:revenue-sol}
f(x,t)\propto t(px + c)^{-1-\frac{1}{p(1-\rho)}}.
\end{equation} 
Rewriting this explicitly as $f(R,t) \propto tR^{-1-\frac{1}{p(1-\rho)}}$ emphasizes the result of a power law distribution of firms in revenue.
This result corresponds with simulation \cite{Axtell2016120, amaral1998}
and empirical data \cite{axtell2001zipf, zhang2009zipf}.
We note that other attempts to quantify this phenomenon have been either empirical, computational, or
statistical in nature; this appears to be the first mechanistic model to naturally generate these dynamics.
\medskip 

\noindent
Figure (\ref{fig:size-powerlaw}) displays the frequency distributions of U.S.\ firms with respect to number of workers from the years 1977 to 2013. 
That a power law fits this data is well-known \cite{axtell2001zipf}. 
As U.S. firms exhibit constant returns to scale \cite{Basu1997returns}, this implies a power law frequency distribution of firms with respect to revenue. 
We note that the average coefficient $\beta_1$ of the log-log fit $\langle \log_{10} f(x) \rangle  = \beta_0 + \beta_1 \log_{10} x$ is approximately equal to 1, implying that $\frac{1}{p(1-\rho)} \approx 0$ in equation (\ref{eq:revenue-sol}). 
The inset of Figure (\ref{fig:size-powerlaw}) displays $\langle f(x, t) \rangle_x$ over the entire date range. 
These data exhibit linear scaling, as predicted by Eq.\ \ref{eq:revenue-sol}, with line of best-fit given by $\langle f(x, t)\rangle_x = -1.98\times 10^{7} + 1.04\times 10^{4}\ t$ ($R^2 = 0.9204$, $p = 8.042 \times 10^{-21}$).
The unpredicted downward trend in $\langle f(x, t) \rangle_x$ in the late 2000s is likely due to unstable conditions in the American economy during this time. 

\subsection{Wealth distribution}
From Eq.\ (\ref{eq:revenue-sol}), we show that the cumulative wealth distribution of firms exhibits power-law scaling. 
Defining the wealth kernel to be $w(x,t) = \pi(x,t)f(x,t)$, where $\pi(x,t)$ are the profits resulting from a sale of $x$ items at time $t$, and imposing a maximum customer base of $x_{\text{max}}$, we have that total system wealth at time $t$ is given by 
\begin{equation}\label{eq:total-wealth}
    W(t) = \int_{t_{\min}}^t \int_{0}^{x_{\text{max}}} w(x',t')\ dx'\ dt'.
\end{equation}
The functional form of $\pi(x,t)$ is dependent on the functional form of firms' cost function $C(x,t)$. 
Suppose that firms face identical weakly quadratic costs $C(x,t) = x_0 +  C'(x_0)(x - x_0) +
\frac{1}{2}C''(x_0)(x
- x_0)^2 + \mathcal{O}\left((x - x_0)^3 \right) \sim cx + \varepsilon x^2$, where we assume $0 < \varepsilon \ll 1$.
(We choose $\eps$ in this range so as to enforce the first-order condition that $\frac{d\pi}{dx} = 0$ has a solution in $\mathbb{R}_+$.)
Then, denoting the total wealth of firms with $x$ or more customers at time $t$ by $W_{\geq}(x,t)$ and letting $p_{\text{net}} = p - c$, the above equation becomes 
\begin{align}
    W_{\geq}(x,t) &\sim \int_{t_{\min}}^t \int_{x}^{x_{\text{max}}} (p_{\text{net}}x' - \varepsilon x'^2)t'x'^{-1-\frac{1}{p(1-\rho)}}\ dx'\ dt' \notag \\
                  &\sim p_{\text{net}}\int_{t_{\min}}^t t' dt' \int_x^{x_{\text{max}}}x'^{-\frac{1}{p(1-\rho)}} dx' \notag \\
&\propto p_{\text{net}}t^2\Big(x_{\text{max}}^{1-\frac{1}{p(1-\rho)}} - x^{1-\frac{1}{p(1-\rho)}}\Big)
\end{align}
Thus the wealth fraction belonging to firms with customer base greater than or equal to $x$ at time $t$, denoted $W_{\text{frac}}(x) = W_{\geq}(x,t) / W(t)$, displays power law scaling as a direct result of the preferential attachment process:
\begin{equation}\label{eq:wfrac}
W_{\text{frac}}(x) \simeq 1 - c_0x^{1-\frac{1}{p(1-\rho)}},
\end{equation}
where $c_0 \approx \Big(x_{\text{max}}^{1 - \frac{1}{p(1-\rho)}}\Big)^{-1}$. 
\begin{figure}
  \centering
  \includegraphics[width=\columnwidth]{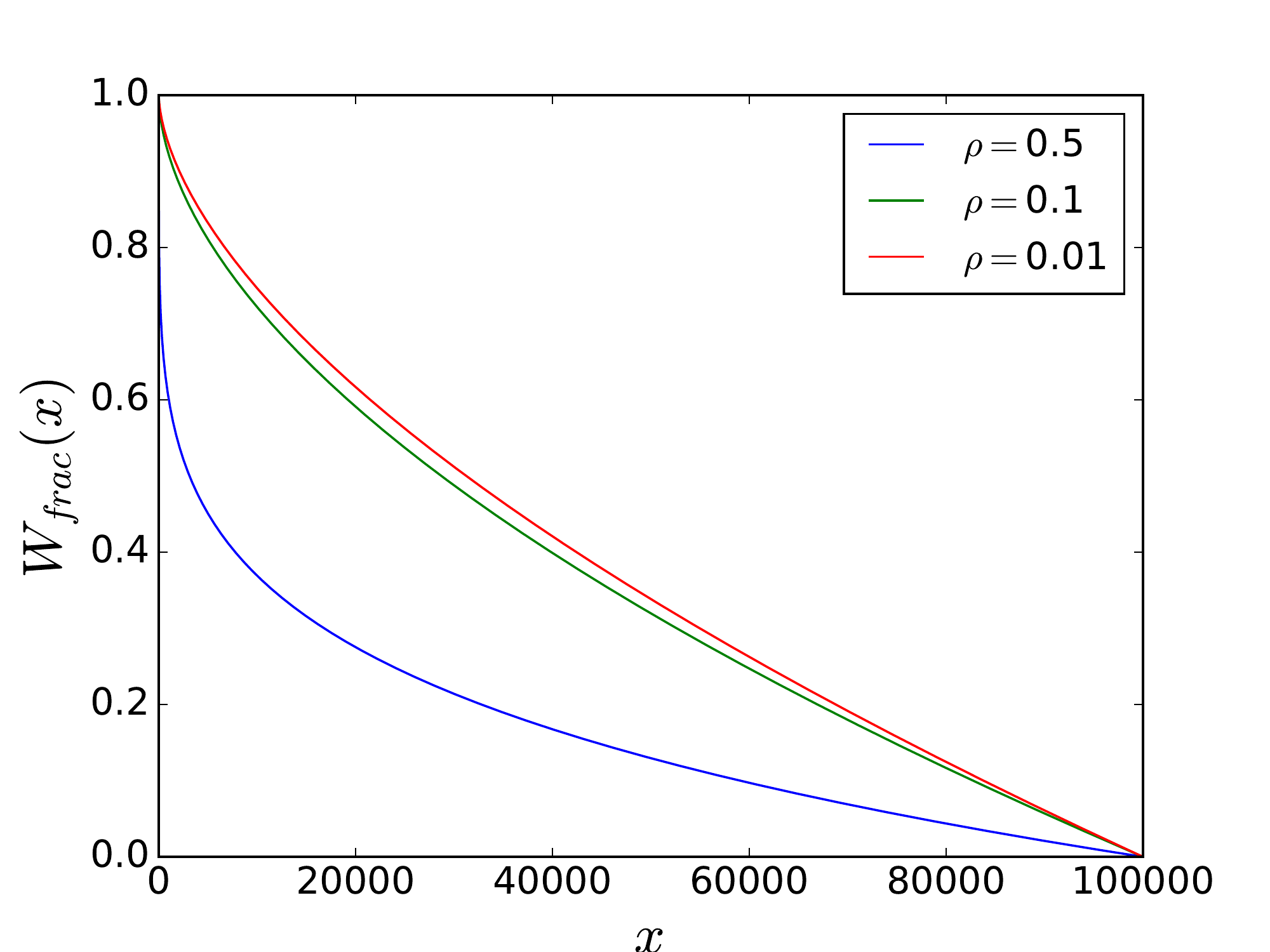}
  \caption{ 
  The fraction of wealth belonging to firms with customer bases greater than or equal to $x$ for $\rho =0.5,\ 0.1,$ and 0.01. 
  We set the price level $p = 2.5$ for ease of visualization. 
  }
  \label{fig:wfrac-plot}
\end{figure}
Eq. \ref{eq:wfrac} is shown in Figure \ref{fig:wfrac-plot} for several values of $\rho$. 

\section*{Concluding remarks and further extensions}

In sum, we have shown how Simon's model, and preferential attachment models more generally, may be extended to the continuum for ease of use as mean-field approximations to stochastic processes. 
We have developed Simon's model in a continuum, expanded upon it by introducing an arbitrary growth kernel $r(x)$ and a time-variant innovation rate $\rho(t)$, and solved the model, discussing the cases in which the general solution satisfies the boundary conditions of the PDE. 
We are able to find explicit solutions with various growth kernels $r(x)$.
Noting that preferential attachment processes may operate in more than one dimension, we allowed the model to have an arbitrary number of dimensions and solved it there. 
Finally, we applied the model to the case of consumer accumulation to firms, to demonstrate a theoretical derivation of the power law distribution of firms by revenue that is observed empirically.\\
~\\
Further extensions to this model could consider the case in which, as treated above, market entrants create their own firms with probability $\rho$. 
It might be that market entrants create not a single firm, but multiple firms, or that, in times of economic crisis, firms are removed from the marketplace with probability $q$.
The model could then be described by
\begin{equation}\label{eq:future-with-g}
\frac{\partial f}{\partial t} = -\frac{1 - \rho}{t}\frac{\partial}{\partial x}\left( r(x)f \right) + q\cdot
	g\left( f(x, t), x, t \right).
\end{equation}
Other models could also incorporate past information about the state of the market or a threshold condition via an equation of the form
\begin{equation}\label{eq:future-with-int}
\begin{aligned}
\frac{\partial f}{\partial t} &= -\frac{1 - \rho}{t}\frac{\partial}{\partial x}\left( r(x)f \right) +\\
&\qquad \int_{x_1}^{x_2}\int_{t_1}^{t_2}h(f(x - x', t - t'), x', t') dt'dx'.
\end{aligned}
\end{equation}
As a concluding example of the possible further generalizations,
let us consider the problem presented in Eq.\ \ref{eq:pde-gen} in a non-asymptotic setting; 
consumers form their own firm at rate $\rho(t)$ with initial firm intensity given by $\mathcal{I}(x-x_0)$.
The model is thus governed by 
\begin{equation}
	\frac{\partial f}{\partial t} = -\frac{1 - \rho(t)}{t}\frac{\partial}{\partial x}\left(r(x)f\right) 
	+\rho(t) \mathcal{I}(x-x_0).
\end{equation}
Again using the method of characteristics, we solve the equations
\begin{align}
	\frac{dx}{dt} &= \frac{1 - \rho(t)}{t}r(x) \\
	\frac{df}{dx} &= -\frac{r'(x)}{r(x)}f(x,t) + \frac{t \rho(t)}{1-\rho(t)}\mathcal{I}(x-x_0)
\end{align}
Solving and again setting the constants of integration $A$ and $B$ to $B = F(A)$, the general solution is given by 
\begin{equation}
	\begin{aligned}
		f(x,t) &= \frac{1}{r(x)}\Bigg[F\left(\int\frac{1-\rho(t)}{t}dt - \int \frac{dx}{r(x)} \right)\\
		&\qquad\quad +  \frac{t \rho(t)}{1 - \rho(t)}\int \mathcal{I}(x-x_0)\ dx \Bigg]
	\end{aligned}
\end{equation}




%

\clearpage

\end{document}